\documentclass[fleqn,twoside]{article}
\usepackage{espcrc2}

\usepackage{graphicx}
\usepackage[figuresright]{rotating}

\newcommand{\AmS}{{\protect\the\textfont2
  A\kern-.1667em\lower.5ex\hbox{M}\kern-.125emS}}

\title{BAIKAL experiment: status report}

\author{V.Balkanov\address[INR]{Institute for Nuclear Research, Moscow, Russia},
I.Belolaptikov$^g$,
N.Budnev\address[IGU]{Irkutsk State University, Irkutsk,Russia},
L.Bezrukov\addressmark[INR],
A.Chensky\addressmark[IGU],
I.Danilchenko\addressmark[INR],
Zh.-A.Dzhilkibaev\addressmark[INR],
G.Domogatsky\addressmark[INR],
S.Fialkovsky$^e$, 
O.Gaponenko\addressmark[INR],
O.Gress\addressmark[IGU],
T.Gress\addressmark[IGU],
R.Il'yasov\addressmark[INR],
D.Kiss$^i (\dag)$,
A.Klabukov\addressmark[INR],
S.Klimushin\addressmark[INR],
K.Konischev\addressmark[INR],
A.Koshechkin\addressmark[INR],
L.Kuzmichev\address[MSU]{Skobeltsyn Institute of Nuclear Physics  MSU, Moscow, Russia},
V.Kulepov\address[NOV]{Nizhni Novgorod State Technical University},
Vy.Kuznetzov\addressmark[INR],
B.Lubsandorzhiev\addressmark[INR],
R.Mirgazov\addressmark[IGU],
N.Moseiko\addressmark[MSU],
M.Milenin\addressmark[NOV],
E.Osipova\addressmark[MSU],
A.Pavlov\addressmark[IGU],
L.Pan'kov\addressmark[IGU],
A.Panfilov\addressmark[INR],
E.Pliskovsky\address[JINR]{Joint Institute for Nuclear Research, Dubna, Russia},
A.Klimov\address[KUR]{Kurchatov Institute, Moscow, Russia},
P.Pokhil\addressmark[INR],
V.Polecshuk\addressmark[INR],
E.Popova\addressmark[MSU],
V.Prosin\addressmark[MSU],
M.Rosanov\address[LEN]{St.Peterburg State Marine University, St.Peterburg, Russia},
V.Rubtzov\addressmark[IGU],
Y.Semeney\addressmark[IGU],
Ch.Spiering\address[DESY]{DESY--Zeuthen, Zeuthen, Germany},
O.Streicher\addressmark[DESY],
B.Tarashanky\addressmark[IGU],
T.Thon\addressmark[DESY],
G.Toth\address[KFKI]{KFKI, Budapest, Hungary},,
R.Vasiliev\addressmark[INR],
R.Wischnewski\addressmark[DESY],
I.Yashin\addressmark[MSU],
V.Zhukov\addressmark[INR]
}
       
\begin{document}

\begin{abstract}
We review the present status of the Baikal Neutrino Project and present the
results obtained with the deep underwater neutrino telescope {\it NT-200}
\vspace{1pc}
\end{abstract}

\maketitle

\section{INTRODUCTION}

The Baikal Neutrino Telescope  is deployed in Lake 
Baikal, Siberia, \mbox{3.6 km} from shore at a depth of \mbox{1.1 km}. 
The present stage of the telescope, {\it NT-200} \cite{APP},
was put into operation at April, 1998. 
Results of searches for atmospheric neutrinos, WIMPs and magnetic monopoles
obtained with {\it NT-200} have been presented elsewhere \cite{B2001}.
During three winter seasons, starting with 1998, a Cherenkov EAS array, consisting of four
\mbox{\textit{QUASAR--370}} phototubes was deployed on the ice, just above
the underwater telescope, with the aim to study the angular resolution
of the latter. Analysis of data show that
the angular resolution of underwater telescope for vertical muons
after modest cuts is about 4$^0$.

In the last winter expedition we continued to study the feasibility of acoustic detection
of EAS cores in water with an EAS array and four hydrophones.
During the EAS array life time of 154 hours, almost 2400 showers with
energies above 5 PeV have been recorded.
Coincidence data of the EAS array and hydrophones are presently
analyzed. Also investigations of water parameters have been continued.
Independent measurements of light absorption and scattering have been
performed by BAIKAL and NEMO (A.Capone et al.) groups.
Preliminary results indicate that the two independent sets of optical data 
are in a good agreement.

In the course of the last expedition we also lowered a special
string with instruments for diverse goals, in particular to
measure the group velocity of light in water at two
different wavelengths and to test a two-channel optical module 
and a calibration  light beacon.

Below, we present new results of a search for a diffuse 
high energy neutrino
flux with the neutrino telescope {\it NT-200}.

\section{A SEARCH FOR HIGH ENERGY NEUTRINOS}

The used search strategy for high energy neutrinos relies
on the detection of the Cherenkov light emitted by the electro-magnetic 
and (or) hadronic particle cascades and high energy muons
produced at the neutrino interaction
vertex in a large volume around the neutrino telescope.
A cut is applied which accepts only time patterns 
corresponding to upward traveling light signals. 

Within the 234 days of the detector livetime, $1.67 \cdot 10^8$ events
with $N_{hit} \ge 4$ have been selected. 
For this analysis we used events with N$_{hit}>$10. 

%%%%%%%%%%%%%%%%%%%%%%%%%%%%%%%%%%%%%%%%%%%%%%%%%%%%%%%%%%%%
\begin{figure}[htbp]
\vspace{-0.6cm}
 \includegraphics[width=5.5cm]{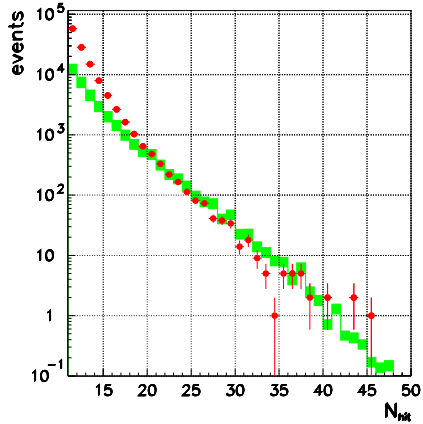}
%\mbox{\epsfig{figure=vfig2.eps,width=6.0cm}}
\vspace{-0.7cm}
\caption{Distribution of hit channel multiplicity; dots - experiment,
hatched boxes - expectation from brems and hadronic showers
produced by atmospheric muons.}
%\label{fig2}
\end{figure}
%%%%%%%%%%%%%%%%%%%%%%%%%%%%%%%%%%%%%%%%%%%%%%%%%%%%%%%%%%%%

Fig.1 shows the N$_{hit}$ distribution for experiment (dots)
as well as the one expected for the background from brems- and
hadronic showers produced by atmospheric muons (boxes).
The experimental distribution is consistent with the 
%theoretical
background
expectation for N$_{hit}>$18. For lower N$_{hit}$ values the 
contribution of  atmospheric muons close to horizon 
as well as low energy showers from $e^+e^-$ pair production
become important.
%The highest multiplicity of hit channels experimentally observed is 
%$N_{hit}^{max}=45$ (one event).
No statistically significant excess over background
expectation from atmospheric
muon induced showers has been observed.
Since no events with $N_{hit}>45$ 
are found in our data we can derive upper limits on the flux of 
high energy neutrinos which would produce events with $N_{hit}>50$. 

The detection volume $V_{eff}$ for neutrino produced events 
with $N_{hit}>$50 which fulfill all trigger 
conditions was calculated as a function of neutrino energy
and zenith angle $\theta$.  $V_{eff}$ rises 
from 2$\cdot$10$^5$ m$^3$ for 10 TeV up to 6$\cdot$10$^6$ m$^3$    
for $10^4$ TeV and significantly exceeds the geometrical volume
$V_{g} \approx$ 10$^5$ m$^3$ of {\it NT-200}. 

Given an $E^{-2}$ behaviour of the neutrino spectrum and a
flavor ratio 
$(\nu_e+\tilde{\nu_e}):(\nu_{\mu}+\tilde{\nu_{\mu}})=1:2$,  
the combined 90\% C.L. upper limit obtained with the Baikal 
neutrino telescopes 
{\it NT-200} (234 days) and {\it NT-96} \cite{APP3} (70 days) 
is:

\begin{equation}
\Phi_{(\nu_e+\tilde{\nu_e})}E^2<(1.3 \div 1.9)\cdot10^{-6} 
\mbox{cm}^{-2}\mbox{s}^{-1}\mbox{sr}^{-1}\mbox{GeV}
\end{equation}
%where the upper value refers to the conservative limit on light
%scattering in the Baikal water.
where the upper value allows for the strongest light scattering
observed over many seasons.

%%%%%%%%%%%%%%%%%%%%%%%%%%%%%%%%%%%%%%%%%%%%%%%%%%%%%%%%%%%%
\begin{figure}[htb]
\vspace{-0.7cm}
%\mbox{\epsfig{figure=vfig7.eps,width=6.5cm}}
\includegraphics[width=6.3cm]{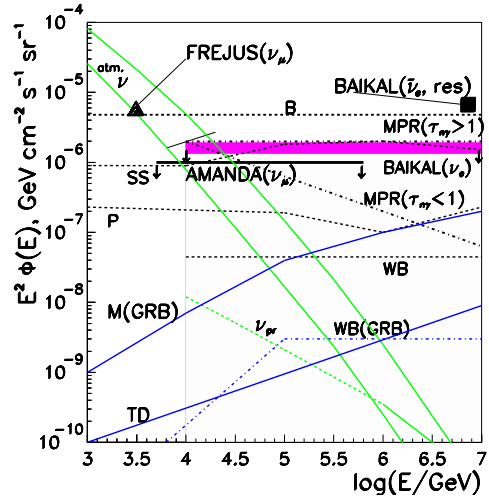}
\vspace{-0.7cm}
\caption{Experimental upper limits on the neutrino fluxes as well as flux 
predictions in different models of neutrino sources (see text).
}
\end{figure}
%%%%%%%%%%%%%%%%%%%%%%%%%%%%%%%%%%%%%%%%%%%%%%%%%%%%%%%%%%%%

Fig.2 shows the upper limits on the isotropic diffuse neutrino
flux obtained by $\,$ BAIKAL $\,$(this work), 
\mbox{AMANDA} \cite{AMANDA2}
and FREJUS \cite{FREJUS} (triangle) 
as well as the atmospheric conventional neutrino \mbox{fluxes \cite{VOL}} 
from horizontal and vertical directions (upper and lower curves,
respectively) and atmospheric prompt neutrino flux  
\cite{PROMT} (curve labeled $\nu_{pr}$). 
Also shown is the model  
independent upper limit on the diffuse high energy neutrino flux
obtained by Berezinsky \cite{Ber3} (curve labeled 'B'),
%with the energy density of the diffuse X- and gamma-radiation 
%$\omega_x \leq 2 \cdot 10^{-6}$ eV cm$^{-3}$ (as follows from
%EGRET data \cite{EGRET}),  
and predictions for diffuse neutrino fluxes
from Stecker and Salamon  \cite{SS} 
('SS') $\,$ and 
$\,$ Protheroe \cite{P} ('P').
Curves labeled 'MPR' and 'WB' show the upper bounds obtained by 
Mannheim et al.
\cite{P98} as well as the upper bound obtained
by Waxman and Bahcall \cite{WB1}, respectively.
Curves labeled 'M(GRB)' and 'WB(GRB)' present the upper bounds for diffuse
neutrino flux from GRBs derived by Mannheim \cite{MANNHEIM} and
Waxman and Bahcall \cite{WB2}. 
The curve labeled 'TD' shows the prediction for neutrino flux from topological
defects due to specific top-down scenario BHS1 \cite{BHS1}.

Our combined 90\% C.L. limit at the W - resonance energy is:

\begin{equation}
\frac{d\Phi_{\bar{\nu}}}{dE_{\bar{\nu}}} \leq (1.4 \div 1.9) \times 
10^{-19} 
\mbox{cm}^{-2}\mbox{s}^{-1}\mbox{sr}^{-1}\mbox{GeV}^{-1}
\end{equation}
and is given by the rectangle in Fig.2.

%The upper limits (1), (2) 
%obtained for the diffuse E$^{-2}$ ($\nu_e+\tilde{\nu_e}$) flux and the model 
%independent $\tilde{\nu_e}$ flux at resonant energy 6.3$\cdot$10$^6$GeV 
%are the most stringent ones at present. 

\section{CONCLUSION}

The neutrino telescope {\it NT-200}  is 
taking data since April 1998. 
It performs investigations of atmospheric muons and neutrinos, 
and searches for WIMPs, magnetic monopoles
and extraterrestrial high energy neutrinos. 
%%%%%%%%%%%%%%%%%%%%%%%%%%%%%%%%%%%%%%%%%%%%%%%%%%%%%%%%%%%%
\begin{figure}[htb]
%\vspace{-0.5cm}
%\mbox{\epsfig{figure=enh_nt_col.eps,width=4.5cm}}
\includegraphics[width=5.0cm,height=6.3cm]{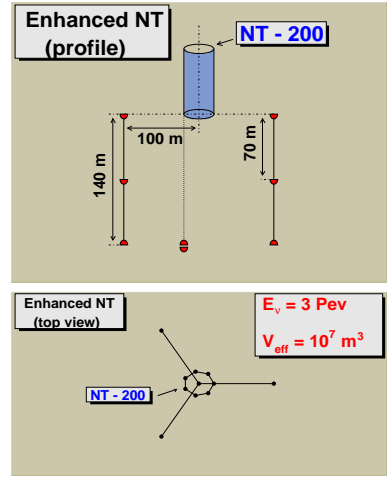}
\vspace{-0.7cm}
\caption{Sketch of {\it NT-200}+.
}
\end{figure}
%%%%%%%%%%%%%%%%%%%%%%%%%%%%%%%%%%%%%%%%%%%%%%%%%%%%%%%%%%%%

In the next 2 years we plan to 
increase the sensitivity to diffuse fluxes by a factor of four.
With a moderate upgrade of only 22 optical modules
at three additionals strings we would reach
a sensitivity of 
$\Phi_{\nu}E^2 \leq 3\cdot10^{-7}$cm$^{-2}$s$^{-1}$sr$^{-1}$GeV.
This upgrade towards a 10\,Mton detector {\it NT-200}+  
%NT-200M 
is 
sketched in fig.3.

This work was supported by the Russian Ministry of Research 
(contract \mbox{\sf 102-11(00)-p}), the German 
Ministry of Education and Research and the Russian Fund of Basic 
Research (grants  \mbox{\sf 99-02-18373a}, 
\mbox{\sf 01-02-31013},\mbox{\sf 00-15-96794} and \mbox{\sf 01-02-17227}),
and by the Russian Federal Program ``Integration'' (project no. 346).

\end{document}